\documentclass[aip,apl,showpacs,showkeys,preprint,superscriptaddress,preprintnumbers,amsmath,amssymb]{revtex4-1}
\usepackage{graphicx}
\usepackage{dcolumn}
\usepackage{bm}
\usepackage{xspace} 
\usepackage{dsfont} 
\usepackage{psfrag}
\usepackage{bbm}
\usepackage{hyperref}
\usepackage{float}
\usepackage[caption=false]{subfig}
\usepackage[usenames]{color}
\usepackage{textcomp}
\usepackage{epstopdf}
\usepackage{wasysym}
\usepackage{bibunits}

\defaultbibliographystyle{apsrev4-1}
\defaultbibliography{Literature-database}

\begin{document}
\begin{bibunit}
\title{Superconducting Single-Photon Detectors with Enhanced High-Efficiency Bandwidth}
\author{Stephan Krapick}\email[corresponding author: ]{stephan.krapick@gmail.com}
\affiliation{Applied Physics Division, National Institute of Standards and Technology, 325 Broadway, Boulder, CO 80305, USA}
\affiliation{Department of Physics, University of Paderborn, Warburger Str. 100, 33098 Paderborn, Germany}
\author{Marina Hesselberg}
\author{Varun B. Verma}\email[corresponding author: ]{varun.verma@nist.gov}
\author{Igor Vayshenker}
\author{Sae Woo Nam}
\author{Richard P. Mirin}
\affiliation{Applied Physics Division, National Institute of Standards and Technology, 325 Broadway, Boulder, CO 80305, USA}
\date{\today}

\begin{abstract}
We present an alternative approach to the fabrication of highly efficient superconducting nanowire single-photon detectors (SNSPDs) based on tungsten silicide. Using well-established technologies for the deposition of dielectric mirrors and anti-reflection coatings in conjunction with an embedded WSi bilayer photon absorber structure, we fabricated a bandwidth-enhanced detector. It exhibits system detection efficiencies (SDE) higher than $\left(87.1\pm1.3\right)\,\%$ in the range from $1450\,\mathrm{nm}$ to $1640\,\mathrm{nm}$, with a maximum of $\left(92.9\pm1.1\right)\,\%$ at $1515\,\mathrm{nm}$. Our measurements indicate SDE enhancements of up to $\left(18.4\pm1.7\right)\,\%$ over a single-absorber WSi SNSPD. The latter has been optimized for 1550 nm for comparison and exhibits maximum SDE of $\left(93.5\pm1.2\right)\,\%$ at 1555 nm. We emphasize that our technological approach has been tested with, but is not limited to, the wavelengths and absorber material presented here. It could be adapted flexibly for multi-color detector systems from the ultraviolet to the mid-infrared wavelength range. This bears the potential for significant improvements in many current quantum optical experiments and applications as well as for detector commercialization.
\end{abstract}


\maketitle

\section{Introduction}

Single-photon detectors have been implemented in many experiments and applications for the past fifteen years, for example in the fields of quantum communication and astronomy. Among several technological approaches, superconducting nanowire single-photon detectors (SNSPDs) have gained broad attention and applicability since they offer extraordinarily high system detection efficiencies\cite{Marsili2013,Verma2015} (SDE), low timing jitter\cite{Goltsman2001,Dauler2009,Chen2013,Najafi2015a}, fast recovery times\cite{Rosenberg2013,Kerman2013,Verma2014}, and ultra-low intrinsic dark counts\cite{Shibata2015}. Recent fundamental experiments and future applications in quantum optics, for instance loophole-free tests of local realism\cite{Shalm2015}, light-matter-interfaces and quantum memories\cite{Jin2015a,Saglamyurek2016}, quantum teleportation\cite{Takesue2015} and quantum key distribution\cite{Shibata2014}, as well as the verification and characterization of photonic states\cite{Dixon2014,Najafi2015,Weston2016}, would not be possible or might have been much more difficult, if single-photon detector systems with low detection efficiencies and signal-to-noise-ratios were used instead.

Currently available SNSPD systems are often designed and optimized for a specific wavelength\cite{Marsili2013,Zhang2016,Jeannic2016}, exhibiting only a limited bandwidth of their high-efficiency region. This appears contradictory to the fact that the absorption region of the incorporated superconductor materials can reach from near-UV to mid-IR wavelengths\cite{Baek2011}. The main reasons for those designs lie in the low complexity of the fabrication cycle. 

One method to fabricate front-illuminated high-efficiency SNSPDs includes a gold mirror as the backside reflector of a dual-pass optical stack. The highest reported SDE values of $93\,\%$ have been demonstrated for $1550\,\mathrm{nm}$\cite{Marsili2013} and $1064\,\mathrm{nm}$\cite{Jeannic2016} using this fabrication method. However, the detection efficiency can drop significantly for non-optimized wavelengths mainly due to the material dispersion of the optical layer stack. Likewise, the fabrication yield might be limited due to the fact that the optical constants (\textit{i.e.}, real part $n$ and complex part $k$ of the refractive index) of the thin-film metallic reflector are not easy to determine and usually have large uncertainties. Thus, the optical impedances of the periodically structured absorber and the buffer/mirror subsystem are difficult to match reliably. As a consequence the optical absorption efficiency of the SNSPD device is hard to predict with high confidence. 

By contrast, dielectric materials are known for their very low absorption over a broad spectral range from the near-UV to the mid-IR, and their optical constants can be determined conveniently by measuring the wavelength-dependent transmittance and reflectance of thin films with known thicknesses.

In this work we present a technological approach for SNSPDs that provides significant bandwidth enhancements of high system detection efficiencies in the telecommunication E-, S-, C-, L-,  and U-bands. We embedded two WSi nanowires, which are separated by a thin insulating barrier and in the following referred to as a \textit{bilayer absorber}, in an all-dielectric backside-mirror and front-side anti-reflection coating (ARC) \cite{Gaggero2010,Redaelli2016,Zhang2016a,Li2016,Yamashita2016}. With dielectric materials we can predict the performance of the backside mirror more accurately as compared to metal-based reflectors. A bilayer absorber acts as a superconducting nanowire avalanche detector based on the interaction of athermal phonons\cite{Verma2016}. We will show the efficiency improvements related to the bilayer nanowire technology by comparing the SNSPD performance to a single-layer WSi SNSPD.

\section{Device design and fabrication of single- and bilayer SNSPDs}

For both device architectures we fabricated a 13-layer stack of alternating silicon dioxide $\left(\mathrm{SiO}_2\right)$ and amorphous silicon $\left(\alpha\mathrm{Si}\right)$ on a 4 inch silicon handle wafer, using plasma-enhanced chemical vapor deposition (PECVD). This technique offers two major benefits:

1) Both mirror materials are based on silane $\left(\mathrm{SiH}_4\right)$ as a precursor, and they can be deposited in the same fabrication run with excellent reproducibility in terms of thickness and optical constants. The  high physical densities of the deposited materials prevent micropores that could lead to irreversible incorporation of water from the environment. 

2) The refractive index contrast between the two materials is considerably higher than for other combinations of dielectrics, while the material dispersion is low in the wavelength range considered here. Consequently, only a small number of alternating layers is required in order to achieve high mirror reflectance over a broad spectral range.

\begin{figure}
\includegraphics[width=0.75\linewidth]{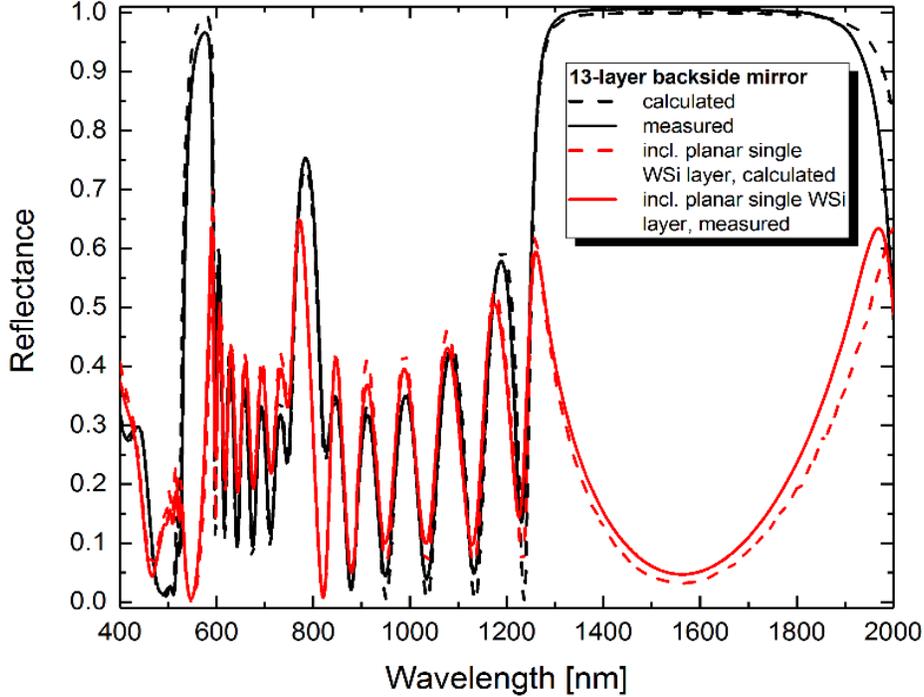}
\caption{\label{fig:01} Measured reflectance curves of the 13-layer dielectric backside mirror without (black) and with (red) a planar single-layer WSi/$\mathrm{SiO}_2$ absorber for comparison. Dashed lines depict simulation results. Maximum reflection of $R\ge0.98$ is achieved in the range of $1350\,\mathrm{nm}\le\lambda\le1800\,\mathrm{nm}$ for the mirror structure, while a minimum reflectance, corresponding to maximized optical absorption efficiency of the planar absorber, can be found at 1562 nm for the mirror/WSi/$\mathrm{SiO}_2$ stack.}
\end{figure}

The deposited individual $\mathrm{SiO}_2$ layers have physical thicknesses of $\tau_{\mathrm{SiO}_2}=(255.2\pm2.5)\,\mathrm{nm}$, while the $\alpha\mathrm{Si}$ thicknesses were $\tau_{\alpha\mathrm{Si}}=(149.1\pm1.2)\,\mathrm{nm}$, both corresponding to a $\lambda/4$ optical stack at $1550\,\mathrm{nm}$. We found that the thermal expansion properties of the two mirror materials allow for damage-free repetitive cooling and warming between room temperature and cryogenic operation temperatures of the SNSPDs. Additionally, a low surface roughness of $r_\mathrm{rms}\le0.5\,\mathrm{nm}$ has been measured for the mirror stack using atomic force micrscopy. Figure \ref{fig:01} depicts the simulated and measured reflectance curves of the planar mirror structure. We infer from the graph that in the region $1350\,\mathrm{nm}\le\lambda\le1800\,\mathrm{nm}$ the reflectance is close to unity ($R_\mathrm{meas}\ge(0.980\pm0.010)$). Therefore, both the transmittance $T$ \textit{and} the absorbance $A$ of the mirror and the handle wafer combined are close to zero. 

The mirror deposition process, identical for both device architectures, was followed by the electron-beam evaporation and lithographic structuring of a Ti/Au electrode layer with thicknesses of $2.0\,\mathrm{nm}$ and $50.0\,\mathrm{nm}$, respectively. 

The single-layer WSi absorber was deposited using magnetron sputtering with nominal thickness of $\tau_{\mathrm{WSi}}=3.5\,\mathrm{nm}$, followed by PECVD deposition of a $\mathrm{SiO}_2$ cap layer with a thickness of $\tau_{\mathrm{cap}}=5.1\,\mathrm{nm}$ in order to prevent significant oxidation of WSi. As an example, we show the reflectance curve of the planar mirror/absorber stack in Fig. \ref{fig:01}, and we see that at around 1562 nm the reflectance approaches a minimum of $R_\mathrm{abs}=(0.047\pm0.010)$, corresponding to an absorption efficiency $\eta_\mathrm{abs}=(95.3\pm1.0)\,\%$. While the wavelength of minimum reflectance fits very well to the simulation performed with rigorous coupled-wave analysis (RCWA\cite{Moharam1981,Li2016}), the reflectance magnitude differs by around $2\,\%$. We suspect this and also the difference in the low-reflectance bandwidth to originate from slight differences in the optical constants and/or thickness values  used in our simulation. The absolute intensity calibration of the spectrophotometer used may also contribute to the discrepancies between measurement and the simulations.

We patterned the single-layer absorber with meandering nanowires in a circular area of $\diameter=20\,\mu\mathrm{m}$, using electron-beam lithography (EBL) and subsequent $\mathrm{SF}_6$ dry etching. The wire widths were $w_\mathrm{SL}=140\,\mathrm{nm}$ and the gaps around $g_\mathrm{SL}=80\,\mathrm{nm}$. Since the resulting non-unit fill factor leads to a reduction of the optical absorbance (also referred to as optical absorption efficiency $\eta_\mathrm{abs}$), we deposited a 3-layer ARC of $\mathrm{SiO}_2/\alpha\mathrm{Si}/\mathrm{SiO}_2$ with nominal thicknesses of $223\,\mathrm{nm}/173\,\mathrm{nm}/154\,\mathrm{nm}$ using PECVD.

\begin{figure}
\includegraphics[width=0.75\linewidth]{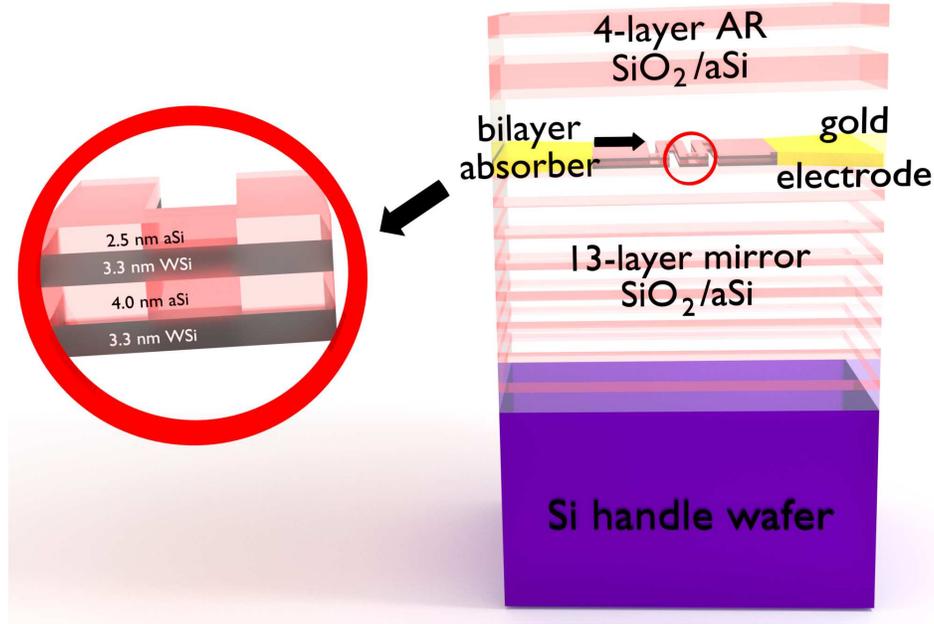}
\caption{\label{fig:02} Schematic of the bilayer device: A planar 13-layer mirror stack comprising $\mathrm{SiO}_2$ and amorphous silicon is deposited on a silicon handle wafer. The meandering nanowire absorber consists of a WSi/$\alpha$Si/WSi/$\alpha$Si structure (bottom to top), which is followed by a 4-layer $\mathrm{SiO}_2$/$\alpha$Si anti-reflection coating.}
\end{figure}

A schematic of the optical stack of our bilayer device is depicted in Figure \ref{fig:02}.  In contrast with the single-layer architecture, on top of an identically deposited dielectric mirror and gold electrodes, the bilayer device uses a structure of WSi/$\alpha\mathrm{Si}$/WSi/$\alpha\mathrm{Si}$ with thicknesses of $3.3\,\mathrm{nm}/4.0\,\mathrm{nm}/3.3\,\mathrm{nm}/2.5\,\mathrm{nm}$ fabricated by magnetron sputtering. While the first layer of amorphous silicon acts as an electrical insulator, the second one is again a protective capping layer. We patterned the bilayer absorber into $24\,\mu\mathrm{m}\times24\,\mu\mathrm{m}$-wide squares as meandering nanowires with wire widths of $w_\mathrm{BL}\approx110\,\mathrm{nm}$ and gaps of $g_\mathrm{BL}\approx70\,\mathrm{nm}$ using EBL and $\mathrm{SF}_6$ dry etching. In order to guarantee a high optical absorption efficiency over a broad wavelength range, we deposited a 4-layer ARC comprising $\mathrm{SiO}_2/\alpha\mathrm{Si}/\mathrm{SiO}_2/\alpha\mathrm{Si}$ with nominal thicknesses of $222\,\mathrm{nm}/145\,\mathrm{nm}/100\,\mathrm{nm}/45\,\mathrm{nm}$, respectively. We provide simulations of the nanowire grating effect on the absorption efficiency in the supplementary information for both device architectures.

\section{SDE Measurement setup and procedure}

\begin{figure}
\includegraphics[width=0.75\linewidth]{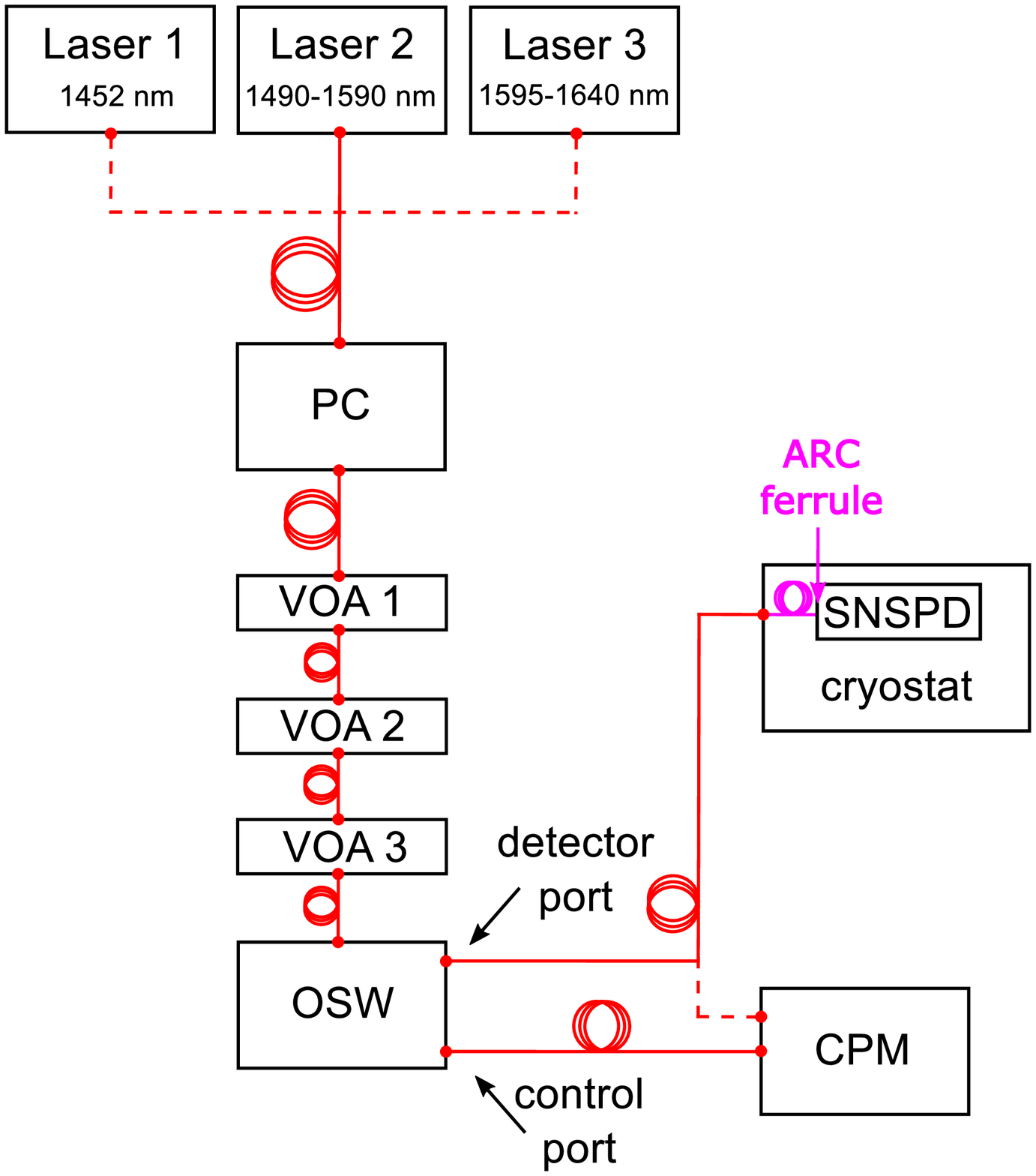}
\caption{\label{fig:03} Schematic of the light path in the SDE measurement setup for setting the photon flux impinging our SNSPDs. All components are remote controlled for identical measurement conditions (see main text for details). PC: polarization controller (automated); VOA: variable optical attenuator; OSW: optical switch; CPM: calibrated control power meter; ARC: anti-reflection coating.}
\end{figure}

We mounted both devices into an adiabatic demagnetization refrigerator (ADR), operated at $T_\mathrm{op}=(280\pm5)$ mK. We implemented the setup for retrieving the SDE as depicted schematically in Fig. \ref{fig:03}. The devices are illuminated using multiple laser sources covering the wavelength range from $1450\,\mathrm{nm}\le\lambda\le1640\,\mathrm{nm}$. We set the power levels so that the power at the lasing wavelength is much greater than the integrated power in the broadband amplified spontaneous emission of the gain media. After having passed an automated polarization controller (PC), the laser light was attenuated by three variable optical attenuators (VOA) in series with overall attenuation levels between $80$ dB and $95$ dB, depending on the wavelength and the available laser power. The light path was further connected to an optical switch (OSW), which routed the photons either to a calibrated control power meter (CPM) or to the cryostat with the devices under test. 

Prior to the actual SDE measurements, we characterized the wavelength-dependent power switching ratios, $r_\mathrm{SW}$, of the OSW at zero attenuation, using the calibrated power meter at both output ports. Inside the ADR we coupled the light to the detectors using anti-reflection coated single-mode optical fibers, aligned to the micromachined SNSPD through commercial zirconia sleeves \cite{Miller2011}, in order to reduce the optical coupling losses. The electrical output pulses of the SNSPDs were amplified and then remotely read out with a counting module.

We determined the wavelength-dependent SDEs as the ratio of the individual photon count rates $PCR$ and the calculated photon fluxes $f^\mathrm{calc}_\mathrm{in}$ at the input of the cryostat:
\begin{equation*}
SDE\left(\lambda\right)=\frac{PCR\left(\lambda\right)}{f^\mathrm{calc}_\mathrm{in}\left(\lambda\right)}\cdot 100\,\%.
\end{equation*}

A detailed description is given in the supplementary information. The uncertainties on the SDE values have been derived in similar procedures to those reported in Refs. \cite{Miller2011,Marsili2013}.

\section{Experimental results and discussion}

\begin{figure}
\includegraphics[width=0.75\linewidth]{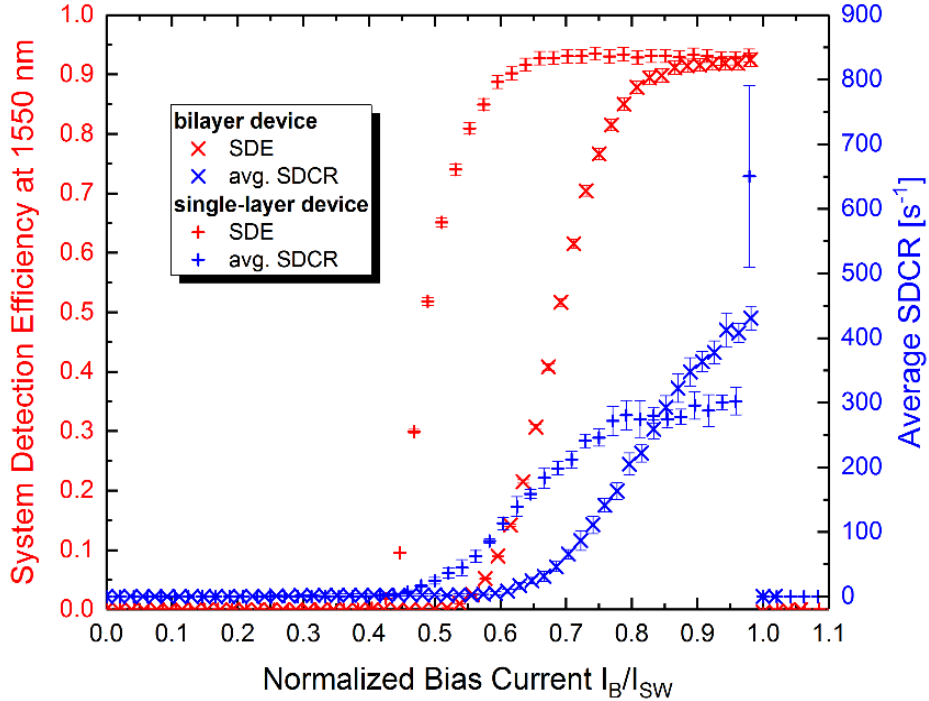}
\caption{\label{fig:04} System detection efficiencies at 1550 nm (red) and system dark count rates (blue) for the single-layer device (+) and the bilayer detector ($\times$), plotted over the normalized bias current $I_\mathrm{B}/I_{\mathrm{SW}}$. While the single-layer detector plateaus at comparably low bias currents, the bilayer device exhibits only a small plateau above $I_\mathrm{B}/I_{\mathrm{SW}}\ge0.9$. The system detection efficiency at $I_\mathrm{B}/I_{\mathrm{SW},BL}=0.98$ is $\left(92.5\pm1.2\right)\,\%$ for the bilayer detector architecture, while for the single-layer device it is $\left(93.2\pm1.2\right)\,\%$ at this wavelength. The system dark count rates of both devices are below 450 $\mathrm{s}^{-1}$ at $I_\mathrm{B}/I_{\mathrm{SW}}\le0.96$. Error bars represent $1\sigma$ standard deviation.}
\end{figure}

For the switching currents, above which the devices exhibit normal conductivity, we measured values of $I_\mathrm{SW,SL}\approx4.8\,\mu$A for the single-layer device and $I_\mathrm{SW,BL}\approx10.6\,\mu$A for the bilayer detector. This difference is consistent with the cross-sectional area of the two nanowire designs. Figure \ref{fig:04} shows the direct comparison of our devices in terms of the average system dark count rate, $\langle SDCR\rangle$, and the inferred system detection efficiency at $1550\,\mathrm{nm}$. Both figures of merit are plotted as dependent on the normalized bias current, $I_\mathrm{B}/I_{\mathrm{SW},i}$. 

The system dark count rates, including scattered residual thermal light within the cryostat, have been determined to be of the same order of magnitude. In particular, the SDCR of the single-layer SNSPD saturated at $\langle SDCR\rangle_\mathrm{SL}=(285\pm25)\,\mathrm{s}^{-1}$, at a bias current of $0.98\cdot I_\mathrm{SW,SL}$, the dark count rate increased significantly and with large fluctuations within multiple measurements. We suspect this to be caused by self-heating hotspots\cite{Kerman2009}, possibly a consequence of the nanowire geometry, but this requires further investigation. For the bilayer device the SDCR exhibited a maximum of $\langle SDCR\rangle_\mathrm{BL}=(431\pm19)\,\mathrm{s}^{-1}$, measured at the highest bias current without switching to normal conductivity, and we did not observe an abnormal increase of the dark count rate here. 

As visualized also in Fig. \ref{fig:04}, both devices exhibit saturated internal detection efficiencies, although the plateaus start at different fractions of the of normalized bias current.  At 1550 nm we found very similar system detection efficiencies for both devices in their plateauing regions. More precisely, at $I_\mathrm{B}/I_{\mathrm{SW},BL}=0.98$ the bilayer device exhibits $SDE_\mathrm{BL}=\left(92.5\pm1.2\right)\,\%$ and the single-layer device shows $SDE_\mathrm{SL}=\left(93.2\pm1.2\right)\,\%$. These values are comparable to the highest reported system detection efficiencies to date. 

\begin{figure}
\includegraphics[width=0.75\linewidth]{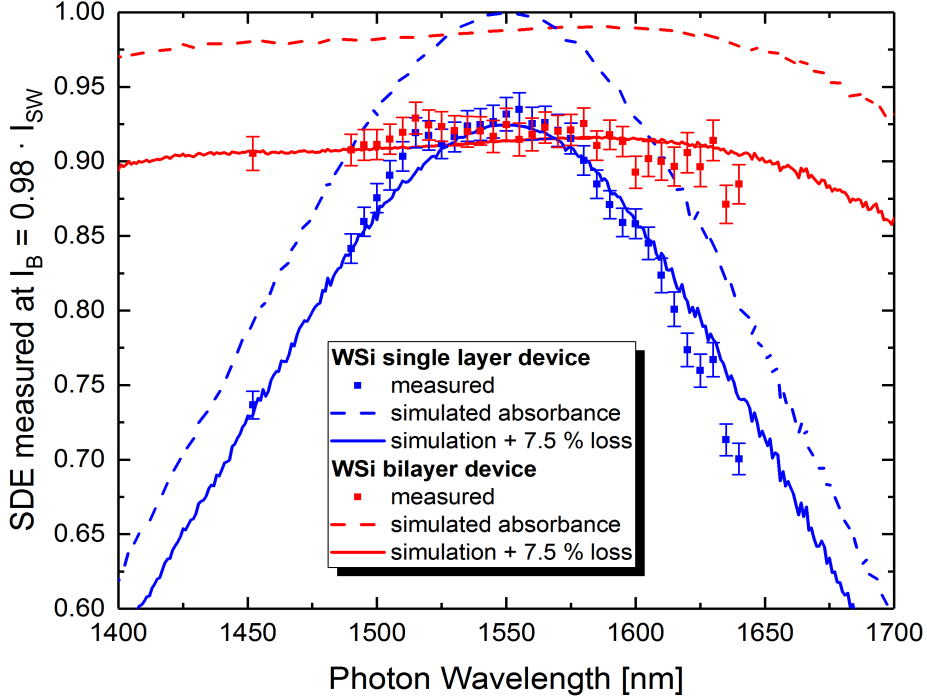}
\caption{\label{fig:05} Direct comparison of the wavelength-dependent average SDEs for the bilayer device (red squares) with the single-layer detector (blue squares). We identify significant improvements of the system detection efficiencies at the boundaries of the available wavelength range up to $\left(18.4\pm1.7\right)\,\%$ in benefit for the bilayer nanowires. At 1550 nm both detectors exhibit SDEs beyond $\ge92\,\%$, and the bilayer device maintains efficiencies of at least $\left(87.1\pm1.3\right)\,\%$ over the spectral range from 1450 nm to 1640 nm. The dashed lines label our RCWA simulation results on the optical absorption efficiency (absorbance), and the solid lines include $7.5\,\%$ loss in the simulation for both devices and match the measured values reasonably well. Note that the waviness of the simulated curves is a result of the implemented dispersion data of the stack materials. Error bars represent $1\sigma$ standard deviation.}
\end{figure}

The SDE wavelength-dependence of the two designs is shown in Fig. \ref{fig:05}. We plotted the system detection efficiencies at $I_\mathrm{B}/I_{\mathrm{SW},BL}=0.98$ versus the input photon wavelength. Clearly, the bilayer device exhibits very high SDE not only at 1550 nm, but over the full wavelength range available in our setup. In particular, we observed a \textbf{minimum} system detection efficiency of $\left(87.1\pm1.3\right)\,\%$ from $1450\,\mathrm{nm}$ to $1640\,\mathrm{nm}$. We found a maximum SDE of $\left(92.9\pm1.1\right)\,\%$ at $1515\,\mathrm{nm}$.

The single-layer detector's SDE peaked as expected at around the center of the telecom C-band. Its maximum SDE of $\left(93.5\pm1.2\right)\,\%$ at 1555 nm is comparable to previously published results based on metal-based backside mirrors, a single $\mathrm{SiO}_2$ buffer layer and a 2-layer $\mathrm{SiO}_2$/$\mathrm{TiO}_2$ ARC\cite{Marsili2013}. We specifically notice the significant drop of the SDE for the single-layer device towards the boundaries of the available wavelength range, \textit{i.e.}, $SDE\left(1450\,\mathrm{nm}\right)=\left(73.7\pm1.0\right)\,\%$ and $SDE\left(1640\,\mathrm{nm}\right)=\left(70.0\pm1.1\right)\,\%$. Comparing both of our devices, these results correspond to wavelength-dependent increases of up to $\left(18.4\pm1.7\right)\,\%$ in favor of the bilayer detector. 

This improvement in bandwidth comes at the cost of an increased polarization dependence. We suspect this to be caused by the lower width-to-thickness-ratio of the bilayer nanowire structure, indicating significant discrepancies in the impedance-matching of the optical stack for different polarizations. From our polarization optimization algorithm (see supplementary information) we deduce that non-optimized photon polarizations lead to significantly lower SDE values, decreased by up to $27\,\%$ to $32\,\%$, depending on the wavelength. For the single-layer device this difference is considerably lower (at maximum around $15\,\%$) in the available wavelength range.

From the graph in Fig. \ref{fig:05}, it is clear that there is a discrepancy of about $7.5\,\%$ between our observed detection efficiency and the predicted absorption efficiency of the detector design.  Although we used a FC/FC mating connector between the source of calibrated photon flux and the input to the cryogenic system, we do not believe all of the $7.5\,\%$ deviation can be explained by the use of the connector. A more likely explanation for the discrepancy is the possibility that the internal quantum efficiency is not $100\,\%$ despite the observation of a saturation in the detection efficiency. The deviation of the measured SDEs from the modeled absorption is very similar to the discrepancy observed in other high efficiency devices based on WSi\cite{Marsili2013}, and it warrants further investigation that is beyond the scope of this paper.

In addition to measurements of the detection efficiency, measurements were performed with a pulsed laser source (mode-locked laser) and an oscilloscope to compare the jitter between the two types of devices. We measured timing jitter (FWHM) of $\tau_\mathrm{jitt,SL} = 200$ ps at a bias currents of $I_\mathrm{B,SL} = 4.7\,\mu$A for the single-layer device, and for the bilayer device, $\tau_\mathrm{jitt,BL} = 159$ ps at a bias current of $I_\mathrm{B,BL}  = 10.2\,\mu$A. The lower timing jitter for the bilayer device is consistent with the higher operating current. However, we found the jitter value for our single-layer device to be higher than for similar single-layer devices made with metallic backside mirrors, using similar readout electronics. One possible explanation could be the difference in RF environment between a metallic backside mirror and a dielectric mirror\cite{Santavicca2016}.

\section{Conclusion and outlook}

We have fabricated, characterized and compared the performances of two WSi-based superconducting nanowire detectors, one as a combination of a bilayer absorber with an all-dielectric optical impedance matching approach and the other including a standard single-layer SNSPD embedded in a similar optical stack. We found that the bilayer detector offers significant enhancements in terms of the system detection efficiency of up to $\left(18.4\pm1.7\right)\,\%$ in the range of $1450\,\mathrm{nm}\le\lambda\le1640\,\mathrm{nm}$ as compared to its single-layer counterpart. While at 1550 nm and $I_\mathrm{B}/I_{\mathrm{SW},BL}=0.98$ the single-layer SNSPD exhibits an SDE of $\left(93.2\pm1.2\right)\,\%$, the bilayer SNSPD shows $\left(92.5\pm1.2\right)\,\%$. Only the bilayer device provides detection efficiencies larger than $\left(87.1\pm1.3\right)\,\%$ over the available wavelength range. This makes it a prospect candidate for quantum optics experiments and applications where high efficiencies over a broad spectral range are crucial for single-photon state detection.

We emphasize that our technological approach for the bilayer absorber might also be conveniently extended or adapted to other superconducting materials and spectral regions, provided the impinging photon energy is higher than the bandgap energy of the absorber films. In particular, we expect SNSPDs based on MoSi,  NbN\cite{Zhang2016a} or NbTiN \cite{EsmaeilZadeh2016} to benefit from analogous technology and exhibiting maximized SDE performance in the near future. 

According to preliminary simulations, it appears feasible to fabricate SNSPDs with high efficiencies at, for example, $(780\pm20)\,\mathrm{nm}$ \textit{and} $(1550\pm40)\,\mathrm{nm}$ simultaneously. Such devices could become an alternative to Silicon-based avalanche photodiodes. Future efforts should also concentrate on improving the timing jitter properties in order to expand the usefulness of our device to time-critical applications such as laser ranging and high-speed quantum cryptography.

\acknowledgments
The authors thank A. E. Lita for her help with the mirror surface roughness measurements and G. C. Hilton for fruitful discussions on the PECVD apparatus.\\

\putbib
\end{bibunit}

\clearpage

\setcounter{figure}{0}
\renewcommand{\thefigure}{S\arabic{figure}}
\renewcommand{\theequation}{S\arabic{equation}}
\renewcommand{\thetable}{S\arabic{table}}
\renewcommand{\bibnumfmt}[1]{[S#1]}
\renewcommand{\citenumfont}[1]{S#1}

\begin{bibunit}
\begin{center}
\textbf{Supplementary information}
\end{center}

In this supplement we provide simulation and measurement results that illustrate the impact of the geometries of single- and bilayer superconducting nanowire single-photon detectors (SNSPDs) on their absorption and system detection efficiencies. In particular, we analyze and compare how the nanowire structuring, in terms of fill factors, influences the bandwidth of the absorption efficiency and the polarization-dependence of the individual device. Additionally, the impact of the deposited anti-reflection coatings (ARC) is studied. Furthermore, we explain the polarization optimization procedure for the system detection efficiency, and we compare the measurement results with our simulations. We conclude that the optical properties of the deposited mirror and ARC materials as well as of the absorber material WSi, in combination with rigorous coupled-wave (RCWA) analysis, predict the optical absorption efficiencies and system detection efficiencies of superconducting nanowire detectors very accurately. We also provide the uncertainty budget for our measurements.

\subsection{Impact of the nanowire structure on the detector absorption efficiency}

\subsubsection{Planar versus patterned absorbers and polarization dependence}

In order to estimate the optical absorption efficiencies (or absorbances, $\eta_\mathrm{abs}$) of the two deposited superconducting nanowire single-photon detector stacks characterized in the main text, we performed simulations based on rigorous coupled-wave analysis (RCWA). In particular, we calculated the wavelength-dependent absorbance after the following fabrication steps: planar nanowire material deposition, nanowire patterning (introduces polarization-dependence), and ARC deposition (manipulates polarization-dependence).

\begin{figure}
\includegraphics[width=0.98\linewidth]{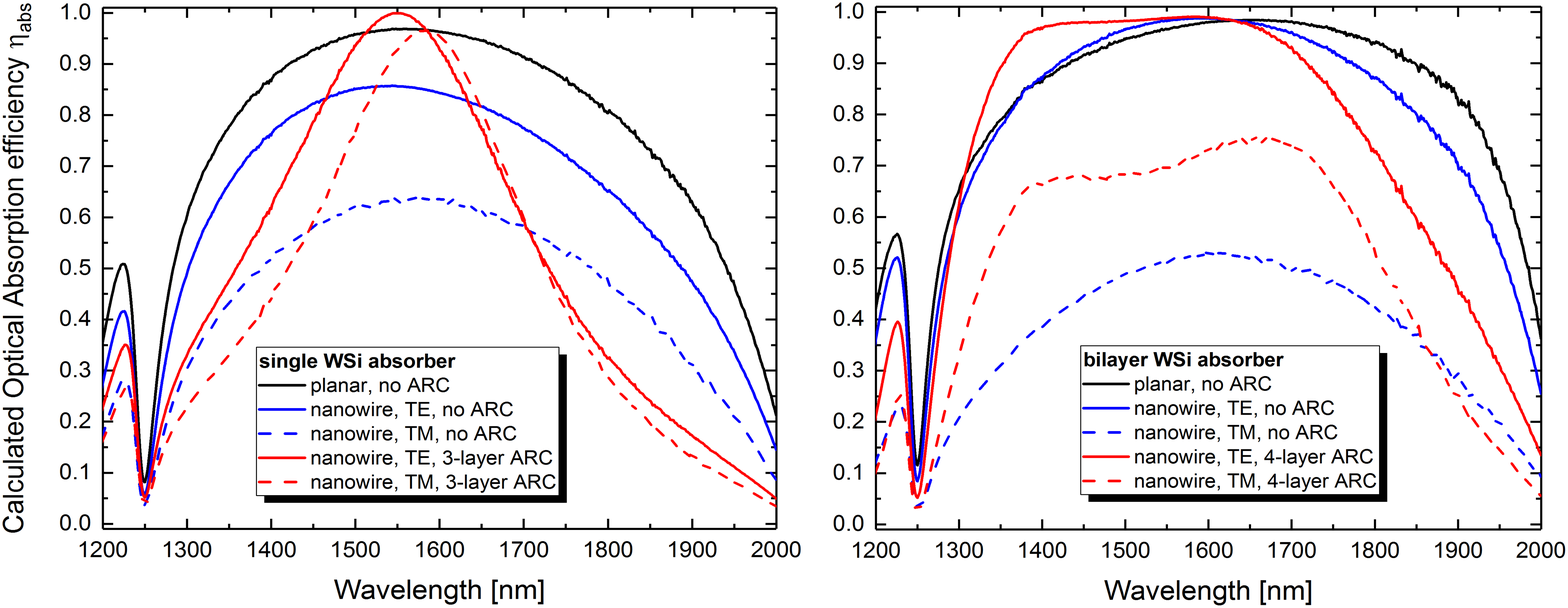}
\caption{\label{fig:s01} Simulated optical absorption efficiencies $\eta_\mathrm{abs}$. The planar absorber on top of the backside mirror exhibits high and broadband $\eta_\mathrm{abs}$ for both design approaches (black lines). a) For the single-layer structure we see that the nanowire patterning reduces the absorption magnitude and peak wavelength for TE-polarized light (blue solid line) by about $12\,\%$, and for TM-polarized light by about $25\,\%$ (blue dashed line), when taking only air as the superstrate into account. After the deposition of the 3-layer anti-reflective coating (red curves), the bandwidth of the recovered high absorption region is significantly narrower. Simultaneously, the polarization dependence is decreased to $\le10\,\%$ at the design wavelength 1550 nm and basically zero at around 1580 nm. b) For the bilayer structure, in contrast to the single-layer counterpart, the nanowire patterning changes the absorption magnitude for TE-polarized light only slightly (blue solid line), but it shifts the spectral bandwidth towards shorter wavelengths. Simultaneously, for TM-polarized light the absorption efficiency is significantly decreased by almost $50\,\%$ (blue dashed line). The deposition of a 4-layer anti-reflective coating helps to maintain a broad bandwidth of the high-absorption region for TE-polarized light between 1400 nm and 1650 nm (red solid line), but for TM-polarized light (red dashed line) the magnitudes are still much lower by about $26\,\%$ to $32\,\%$.}
\end{figure}

Figure \ref{fig:s01}a) shows the simulation results on $\eta_\mathrm{abs}$ for our single-layer device design, accounting for the nominal physical thicknesses as given in the main text. The 
difference between the black curve and the blue solid curve illustrates the reduction of $\eta_\mathrm{abs}$ due to the meandric/periodic absorber patterning. The deposition of a 3-layer ARC allows for the narrow-band recovery of the absorption close to unit efficiency at 1550 nm and for TE-polarized light. The graph also indicates the impact of polarization on the absorption efficiency. The polarization-dependent differences in $\eta_\mathrm{abs}$, caused by a periodic nanowire's grating effect, are less than $10\,\%$ at 1550 nm for the AR-coated device, while exceeding $20\,\%$ for devices with no ARC. Additionally, we found that at 1580 nm the polarization-dependence disappears for the AR-coated devices, as the simulated curves cross each other.

Figure \ref{fig:s01}b) shows the simulation results for the bilayer device. We see that the bilayer absorber patterning does not reduce the maximum achievable absorption efficiency for TE-polarized light, but it causes a peak shift towards shorter wavelengths. This indicates a grating effect of the bilayer structure, caused not only by the dissimilarities of the absorber geometries (cf. Driessen et al. 2009\cite{Driessen2009}), but also by a corresponding effective refractive index being different from the single-layer design. However, for TM-polarized light and without the 4-layer ARC, we see that $\eta_\mathrm{abs}$ is significantly reduced by about $50\,\%$ as compared to TE-polarized light. Our simulations show that the absorbance can be recovered by an ARC close to unit efficiency over a broad spectral range, $1400\,\mathrm{nm}\le\lambda\le1650\,\mathrm{nm}$ for TE modes. For TM modes this could not be achieved in the bilayer nanowires. We anticipate wavelength- and polarization-dependent differences in $\eta_\mathrm{abs}$ between $26\,\%$ and $32\,\%$ from these simulations.

\subsubsection{Impact of the fill factor}

For the patterning of our nanowires we use electron-beam lithography, a technology which may cause inaccuracies in the width and gap of neighboring nanowire sections within the meandric structure. Those local constrictions are typically in the order of a few nanometers, thereby slightly changing the fill factor. Besides the expected impacts on the superconducting properties of the SNSPDs, it is important to know the impact of the electron-beam writing errors on the optical absorption efficiency.

\begin{figure}
\includegraphics[width=0.98\linewidth]{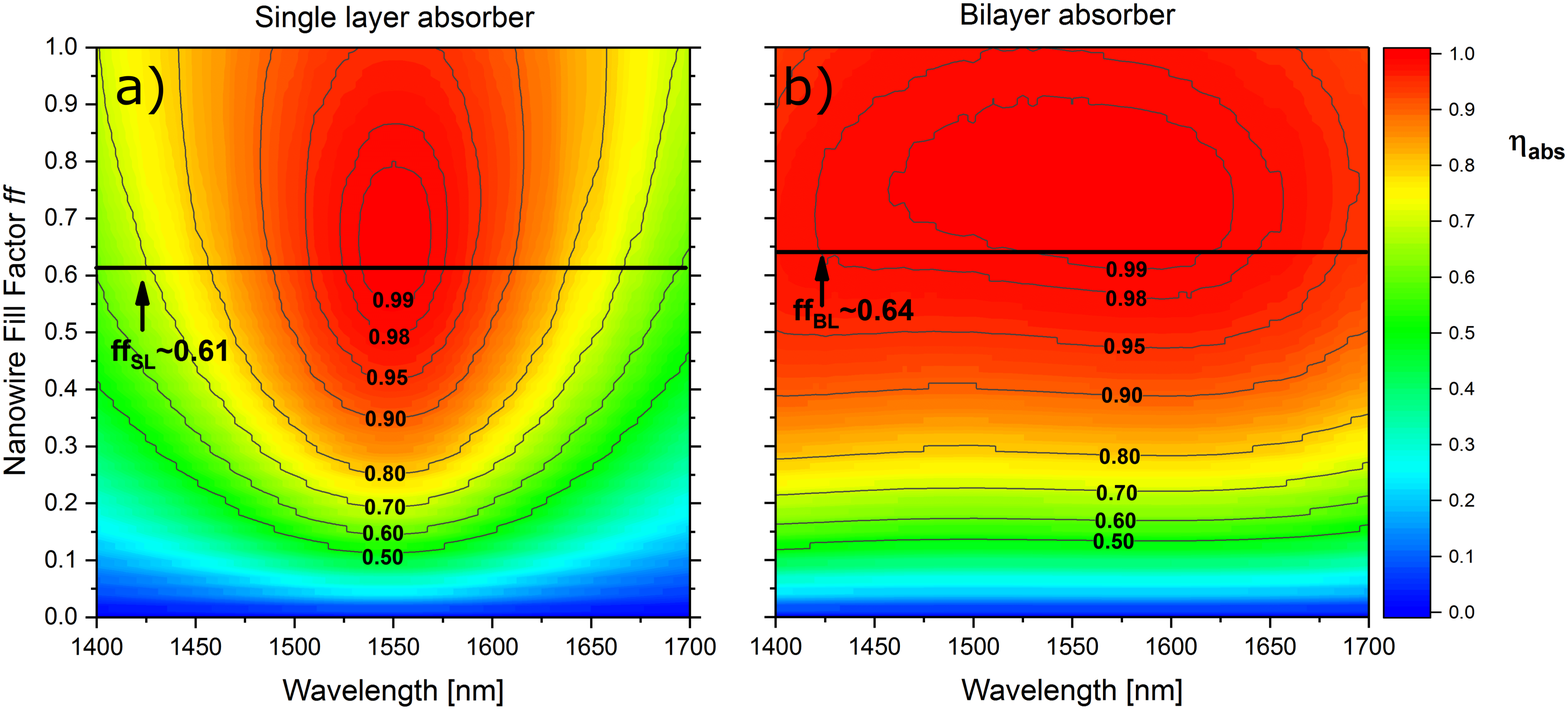}
\caption{\label{fig:s02}  Simulated wavelength- and fill-factor-dependence of the absorption efficiency for TE-polarized photons. a) The single-layer absorber structure exhibits a small region of efficiencies larger than $99\,\%$ around 1550 nm, dropping significantly towards the spectral boundaries of the graph. The fill factor range, where these high absorbances can be achieved, covers the range of $0.55\le ff\le0.78$. b) The simulation shows the spectral enhancement of the high-absorbance region for the optical stack over a broad range of fill factors and wavelengths. Both graphs underline the negligible impact of small changes in the fill factor, \textit{e.g.}, as a consequence of electron beam writing errors, on the absorption efficiency around the design parameter (black bars).}
\end{figure}

We analyzed the influence of the wavelength and the fill factor, given by $ff=w/(w+g)$, where $w$ is the wire width and $g$ is the gap between two nanowire meanders, on the optical absorption efficiency. In our simulations for TE-polarized light, depicted in Fig. \ref{fig:s02}, we used the nominal thicknesses for the optical stack components described in the main text. Obviously the calculated absorption efficiencies for the two different technological approaches show very dissimilar behavior. In particular, the single-layer nanowires exhibit maximum $\eta_\mathrm{abs}$ at 1550 nm when the fill factor is $ff_\mathrm{SL}=0.68$, and it drops rapidly towards longer and shorter wavelengths. The fill factor region for high absorption efficiency ($\eta_\mathrm{abs}\ge0.99$) ranges from $0.55\le ff_\mathrm{SL}\le0.78$, which indicates a reasonable robustness against fabrication inaccuracies, which are very conservatively estimated to be $\pm5\,\%$. We furthermore observe a slight tilt of the iso-efficiency-ellipse in the graph, which means that the wavelength of maximum absorption shifts only slightly with changing fill factor.

Using the nominal thicknesses described in the main text, we find the spectral region of very high absorbance is significantly enhanced for TE-polarized light in the bilayer architecture. We deduce not only that $\eta_\mathrm{abs}\ge0.99$ from 1470 nm to 1650 nm for fill factors $ff_\mathrm{BL}=0.75\pm0.07$, but also that the fill factors have only a minor impact on the absorbance in the range $0.63\le ff_\mathrm{BL}\le0.89$ at C-band telecom wavelengths. So for the bilayer structure we can also infer excellent robustness of the absorbance against nanowire fabrication errors.

\begin{figure}
\includegraphics[width=0.98\linewidth]{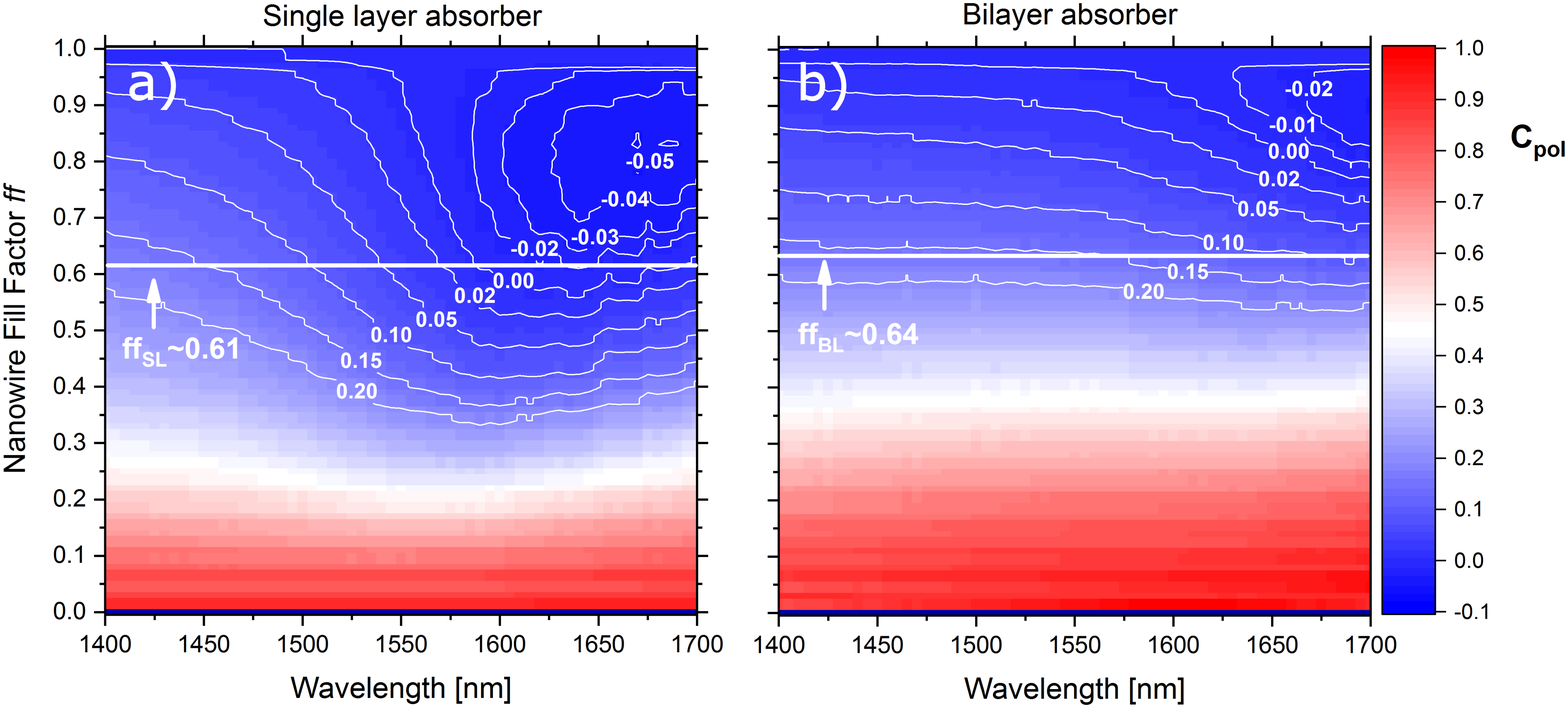}
\caption{\label{fig:s03} Simulated dependence of the polarization-dependent contrast of the absorption efficiency on the wavelength and the fill factor for the two SNSPD architectures. a) The single-layer device exhibits a low polarization-dependent contrast at 1550 nm over a broad range of fill factors above $ff_\mathrm{SL}=0.55$. At given fill factors, the polarization contrast in the absorbance is pronouncedly wavelength-dependent. We find for high fill factors $ff_\mathrm{SL}\ge0.9$ that polarization-independent single-layer devices can be achieved for the telecom C-band. b) The bilayer device shows higher polarization-dependent contrast of the absorbance, but its wavelength-dependence is almost negligible at given fill factors. Increasing the fill factor to $ff_\mathrm{BL}\ge0.9$ leads to significant reduction of the polarization-dependent contrast ($|C_\mathrm{pol}|\le0.02$), which indicates a strong impact of the nanowire geometry (width, gap, and absorber thickness) on the SDE.}
\end{figure}

Similar to the model for TE-modes as mentioned above, we simulated optical response of the structures to TM-polarized light. In order to visualize the impact of the fill factor on the polarization dependence of the two device designs in a given wavelength region, we define a polarization-dependent contrast of the absorbances, $C_\mathrm{pol}$, as follows:
\begin{equation*}
C_\mathrm{pol}=\frac{\eta_\mathrm{abs,TE}-\eta_\mathrm{abs,TM}}{\eta_\mathrm{abs,TE}+\eta_\mathrm{abs,TM}},
\end{equation*}
the values of which must be in the range $-1\le C_\mathrm{pol}\le1$. Here, negative values describe higher nanowire absorbance for TM modes, while positive numbers represent TE-modes to be favored. If $C_\mathrm{pol}=0$, the polarization dependence of the absorption efficiency vanishes, regardless of the absolute values of $\eta_\mathrm{abs}$.

Figure \ref{fig:s03} shows the simulation results on the polarization-dependent efficiency contrast for both device designs. We infer several interesting aspects from both graphs. The single-layer structure in Fig. \ref{fig:s03} a), being optimized for high absorption efficiency at 1550 nm, exhibits a small polarization-dependent efficiency contrast of $C_\mathrm{pol,SL}\le0.05$ at the design wavelength and for fill factors in the range $0.55\le ff_\mathrm{SL}<1.0$. For the design parameters of our fabricated single-layer device, \textit{i.e.}, $ff_\mathrm{SL}\approx0.61$, we identify 1580 nm as the wavelength at which the polarization dependence vanishes. This is in direct correspondence and agreement with the graph in Fig. \ref{fig:s01} a). We also deduce from the simulation that for very small nanowire gaps, where the fill factor exceeds $ff_\mathrm{SL}\ge0.9$, the wavelength for vanishing polarization dependence shifts towards the telecom C-band. Although difficult to fabricate, we believe that single-layer WSi SNSPDs with wire widths of 150 nm, gaps of around 15 nm, and based on a re-optimized ARC, can be candidates for the highly efficient detection of polarization-entangled quantum states at 1550 nm without the need of additional polarization controllers.

In contrast, the simulation in Fig. \ref{fig:s03} b) exhibits a different situation for the bilayer design. Although the polarization-dependent efficiency contrast is almost constant over the considered wavelength range, we find comparably high absolute contrast values which decrease with increasing fill factors. Those results, in combination with the graph in Fig. \ref{fig:s01} b), indicate that the nanowire geometry can be tuned in order to reduce the polarization dependence of the bilayer nanowire efficiency. However, in order to get contrast levels comparable as low as for our fabricated single-layer device, fill factors of $ff_\mathrm{BL}\ge0.89$ would be necessary.

\subsection{Polarization optimization and measured polarization dependence of the SDE on the nanowire architecture}

Within our automated SDE measurement routine, we optimized the polarization of the impinging light such that maximum count rates at a given flux were achieved. In particular, we set the automated polarization controller to the six nominal settings $i$: horizontal (H), vertical (V), diagonal (D), anti-diagonal (A), right-hand-circular (RHC) and left-hand-circular polarization (LHC), and we totalize the photon counts over 10 s of measurement time in order to get the counts $N_i$. From the setting-dependent $N_i$ we calculate the four Stokes parameters according to:

\begin{equation*}
\begin{array}{rcl}
S_0&=&\frac{1}{3}\left(N_\mathrm{H}+N_\mathrm{V}+N_\mathrm{D}+N_\mathrm{A}+N_\mathrm{RHC}+N_\mathrm{LHC}\right),\\
S_1&=&N_\mathrm{H}-N_\mathrm{V},\\
S_2&=&N_\mathrm{D}-N_\mathrm{A},\mathrm{and}\\
S_3&=&N_\mathrm{RHC}-N_\mathrm{LHC}.
\end{array}
\end{equation*}
From these parameters we derived the spherical coordinates, at which the optimal polarization state is set for impinging photons on the SNSPD, \textit{i.e.} where maximum count rates can be expected:
\begin{equation*}
\begin{array}{rcl}
p&=&\frac{\sqrt{S_1^2+S_2^2+S_3^2}}{S_0},\\
\psi&=&\frac{1}{2}\arctan \frac{S_2}{S_1}+k\frac{\pi}{2},\;\mathrm{and}\\
\chi&=&\frac{1}{2}\arctan \frac{S_3}{\sqrt{S_1^2+S_2^2}},
\end{array}
\end{equation*}
where $p$ denotes the polarization degree, $\psi$ is the polarization half angle, and $\chi$ is a measure for the ellipticity of the polarized light. The integer $k$ is chosen such, that for $\frac{1}{2}\arctan \frac{S_2}{S_1}<0$, $k=1$, and else $k=0$.

\begin{figure}
\includegraphics[width=0.75\linewidth]{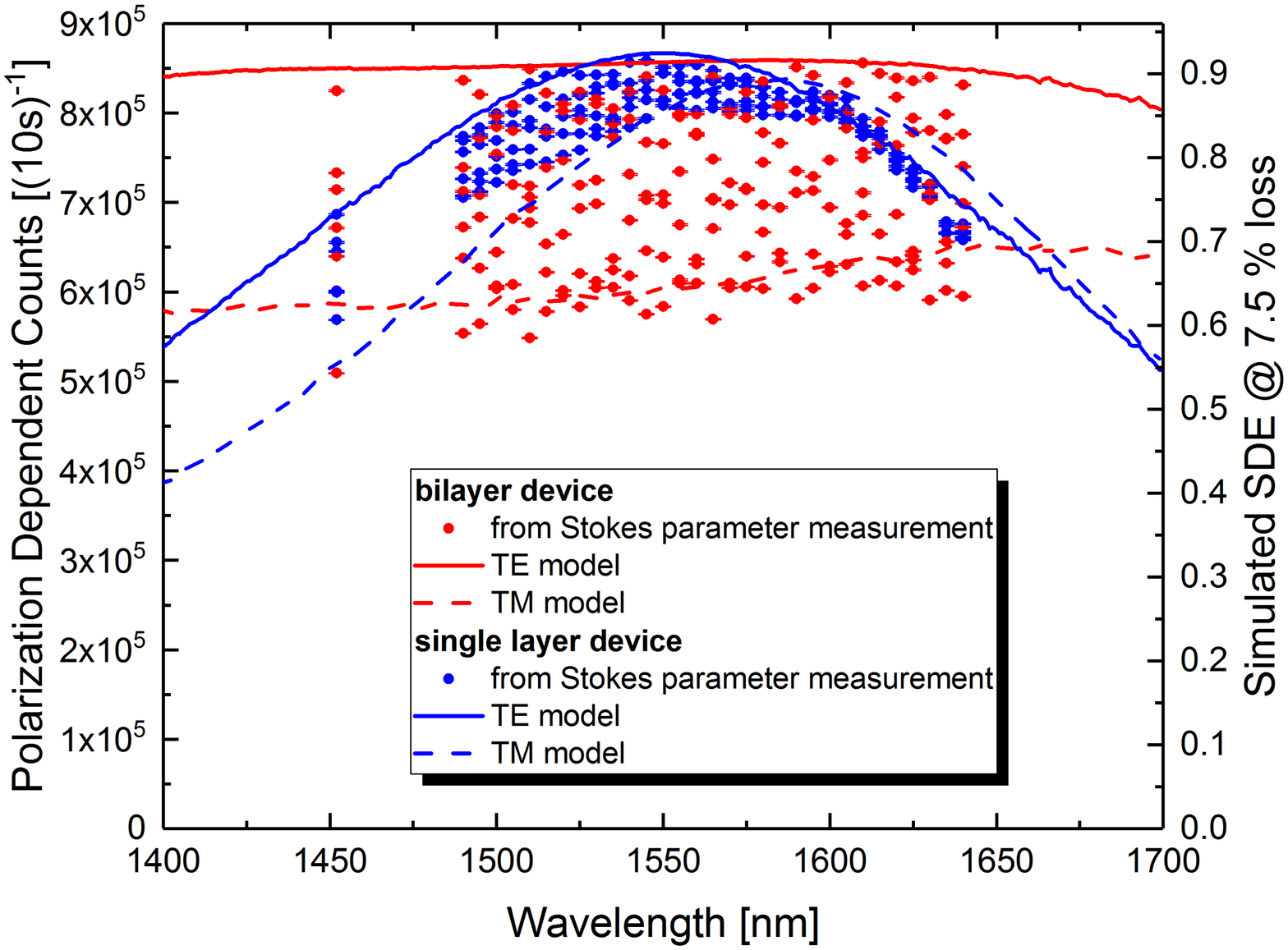}
\caption{\label{fig:s04} Qualitative confirmation of the simulated polarization dependence of the SDE by the scattering of measured counts at different polarization states (see text for detailed explanation). The scattered points refer to the left y-axis, and the lines correspond to the right y-axis.}
\end{figure}

Since this procedure optimizes for the polarization state at which the photon count rates are maximized, we do not explicitly set TE- or TM-polarized states of light. However, for SNSPD devices with a large polarization dependence, the individual totalized counts $N_i$ will scatter around the average value of $S_0/2$, and the Stokes parameters $S_1$, $S_2$ and $S_3$ have large absolute values. At the same time, the scattering will be bound to those two polarization states, at which maximum and minimum system detection efficiency is achieved. In other words, the scattering of the six measured totalized counts $N_i$ at a specific wavelength can provide a qualitative measure of the polarization dependence of the respective SNSPD device. This is depicted in Fig. \ref{fig:s04}, where we plotted the photon counts of the polarization optimization routine versus the wavelength. For comparison, we also plot the theoretical curves of the wavelength-dependent system detection efficiency for TE- and TM-polarized light, including $7.5\,\%$ photon loss. We find reasonably good agreement between measurements and simulations for both the single-layer and also the bilayer SNSPD. In particular, the scattering of $N_i$ for bilayer exceeds the simulated lower bound at TM polarization. Together with our findings in Fig. \ref{fig:s03} b) we suspect this slight discrepancy to be caused by a fill factor in the bilayer SNSPD being actually smaller than the one determined from the scanning electron micrograph. Another explanation for this slight simulation mismatch could be found in the bilayer absorber thicknesses, which might actually be higher than expected from our thickness calibration of the WSi deposition.

\subsection{Uncertainty budget}

The determination of the system detection efficiencies suffers from uncertainties during the measurement process. Here, we provide the $1\sigma$ uncertainty budget for the SDE measurement at 1550 nm. The measurements were taken similar to the procedures reported by Miller et al. \cite{Miller2011} and Marsili et al.\cite{Marsili2013}. The system detection efficiencies have been determined as
\begin{equation}
\label{eq:s01}
SDE=\frac{PCR}{f^\mathrm{real}_\mathrm{input}}\cdot 100\,\%,
\end{equation}
with
\begin{equation}
\label{eq:s02}
PCR=CR-\langle SDCR\rangle
\end{equation}
as the dark-count-subtracted photon count rate, and with
\begin{equation}
\label{eq:s03}
f^\mathrm{real}_\mathrm{input}=\frac{P^\mathrm{real}_\mathrm{CPM}\cdot\alpha_1\cdot\alpha_2\cdot\alpha_3\cdot r_\mathrm{SW}}{\left(1-R_\mathrm{F}\right)\cdot E_\lambda}
\end{equation}
being the real photon flux at the input fiber connector of the cryostat. In Eq. \ref{eq:s03}, $P^\mathrm{real}_\mathrm{CPM}$ denotes the real power at the control power meter, $\alpha_i$ are the individual attenuation factors of the three variable optical attenuators, $r_\mathrm{SW}$ is the wavelength-dependent switching ratio of the optical switch that routes the photon flux to the cryostat input connector, $E_\lambda$ is the photon energy at the wavelength under consideration, and $R_\mathrm{F}$ is the wavelength-dependent Fresnel loss at the uncoated optical fiber connected to the control power meter. We separately characterized the uncertainties for the individual factors of Eq. \ref{eq:s03}. Note that in our calculations we considered the photon energy $E_\mathrm{\lambda}$ and the single-mode optical fiber end-face reflectance $R_\mathrm{F}$ to be known with negligible uncertainty.

The corrected photon count rate, $PCR=CR-\langle SDCR\rangle$, is varying not only with the Poisson standard deviation of the raw count rate $CR$, but it is also influenced by the statistical variation of the system dark count rate $\langle SDCR\rangle$, averaged over multiple dark count measurements. The overall relative uncertainty depends on the applied bias current. We found for 1550 nm and at $98\,\%$ of the switching current that the error-propagated relative uncertainties on the corrected photon count rates are $\sigma_\mathrm{PCR}=0.35\,\%$ for both SNSPDs.

The available real powers of our laser sources are deduced from the actual power readings on the control power meter, $P^\mathrm{read}_\mathrm{CPM}$, by calculating $P^\mathrm{real}_\mathrm{CPM}=\frac{P^\mathrm{read}_\mathrm{CPM}}{CF^\mathrm{cal}_\mathrm{CPM}\cdot CF^\mathrm{NL}_\mathrm{CPM}\cdot CF^\mathrm{RD}_\mathrm{CPM}}$. The wavelength-dependent calibration factor $CF^\mathrm{cal}_\mathrm{CPM}$, the range-dependent nonlinearity correction factor $CF^\mathrm{NL}_\mathrm{CPM}$, and the range discontinuity correction factor $CF^\mathrm{RD}_\mathrm{CPM}$, contribute to the uncertainty budget of an individual power measurement\cite{Vayshenker2006}. Note that we also measured the stability of the laser power. The corresponding uncertainties, derived from the Allan variance, do not significantly contribute to the overall uncertainties and propagated errors. As an example, at 1550 nm, the propagated relative uncertainty of a single power measurement without attenuation is $\Delta P^\mathrm{real}_\mathrm{CPM}(-10\,\mathrm{dBm})=0.39\,\%$, while for attenuated power measurements it is very similar, $\Delta P^\mathrm{real}_\mathrm{CPM}(-40\,\mathrm{dBm})=0.39\,\%$.

The switching ratio has been determined as the ratio of unattenuated powers at the detector port (1) and the control port (2), $r_\mathrm{SW}=P^\mathrm{(1)}_\mathrm{CPM}/P^\mathrm{(2)}_\mathrm{CPM}$. Thus, the uncertainties of the contributing power measurements were propagated, and we find $\Delta { r_\mathrm{SW}}=0.53\,\%$ for the switching ratio at 1550 nm.

For determining the real photon flux at the input of our cryostat, we used the following power calibration sequence prior to each individual measurement of the photon count rate:

1) The offset of the control power meter is determined by measuring the power at zero laser power (beam blocked). 

2) The unattenuated laser power is measured with the control power meter, and we receive $P^\mathrm{real}_\mathrm{CPM,unatt}$ by subtracting the offset from the read power value and by correcting for the calibration/correction factors mentioned above.

3) We measured the powers for all three optical attenuators, given a specific attenuation level and a corresponding range of the control power meter. From these, subtracted by the offset, and from the unattenuated measurements we derived the real attenuation factors, $\alpha_i=\left(P^\mathrm{real}_{\mathrm{CPM,att},i}-P^\mathrm{real}_ {\mathrm{CPM,offset}}\right)/P^\mathrm{real}_\mathrm{CPM,unatt}$, as well as their their relative uncertainties by Gaussian error propagation. At 1550 nm we determined very similar relative uncertainties for all three attenuation factors, \textit{i.e.}, $\Delta \alpha=\Delta \alpha_{i}=0.55\,\%$.

The aforementioned considerations let us conclude that for a single SDE calculation at 1550 nm the \textit{relative} uncertainties are $\Delta SDE\left(1550\,\mathrm{nm}\right)=1.20\,\%$ for both SNSPDs. Note that in the main article we provided the SDE values together with their \textit{absolute} uncertainties, rounded up to one digit and also given in percent. In Table \ref{tab:s01} we have summarized the relative uncertainty contributions on the SDE calculations for 1550 nm and 1640 nm. The wavelength-dependent differences mainly relate to the dissimilar uncertainties on the power meter wavelength calibration factors, on the respective switching ratios, and on the attenuation factors.

\begin{table}[h]
\caption{\label{tab:s01} Uncertainty budget for both SNSPD devices at 1550 nm and 1640 nm.}
\begin{ruledtabular}
\begin{tabular}{lcc}
\textbf{Physical quantity} & \textbf{Bilayer SNSPD} & \textbf{Single-layer SNSPD}\\
& relative uncertainty [\%] & relative uncertainty [\%]\\\hline\hline
$\sigma_\mathrm{PCR} (1550\,\mathrm{nm})$ & 0.35 & 0.35\\
$\sigma_\mathrm{PCR} (1640\,\mathrm{nm})$ & 0.35 & 0.39\\\hline
$\Delta P^\mathrm{real}_\mathrm{CPM}(-10\,\mathrm{dBm},1550\,\mathrm{nm})$ & 0.39 & 0.39\\
$\Delta P^\mathrm{real}_\mathrm{CPM}(-40\,\mathrm{dBm},1550\,\mathrm{nm})$ & 0.39 & 0.39\\
$\Delta P^\mathrm{real}_\mathrm{CPM}(-20\,\mathrm{dBm},1640\,\mathrm{nm})$ & 0.48 & 0.48\\
$\Delta P^\mathrm{real}_\mathrm{CPM}(-50\,\mathrm{dBm},1640\,\mathrm{nm})$ & 0.50 & 0.50\\\hline
$\Delta \alpha_i(-31.6\,\mathrm{dB}, 1550\,\mathrm{nm})$ & 0.55 & 0.55\\
$\Delta \alpha_i(-26.9\,\mathrm{dB}, 1640\,\mathrm{nm})$ & 0.70 & 0.70\\\hline
$\Delta r_\mathrm{SW}(1550\,\mathrm{nm})$ & 0.53 &0.53\\
$\Delta r_\mathrm{SW}(1640\,\mathrm{nm})$ & 0.65 & 0.65\\\hline\hline
$\Delta SDE(1550\,\mathrm{nm})$ & 1.20 & 1.20\\
$\Delta SDE(1640\,\mathrm{nm})$ & 1.49 & 1.50\\
\end{tabular}
\end{ruledtabular}
\end{table}

\putbib
\end{bibunit}

\end{document}